\documentstyle[12pt,fleqn,aaspp4]{article}
\newcommand{\beq}{\begin{equation}}
\newcommand{\eeq}{\end{equation}}

\def\yr{{\rm yr}}

\def\dol{{d_{\rm ol}}}
\def\dls{{d_{\rm ls}}}
\def\dos{{d_{\rm os}}}
\def\deg{{\rm deg}}
\def\psf{{\rm psf}}
\def\texp{{t_{\rm exp}}}

\def\min{{\rm min}}

\def\ccd{{\rm ccd}}
\def\sky{{\rm sky}}
\def\lmc{{\rm LMC}}
\def\E{{\rm E}}

\def\tot{{\rm tot}}

\def\day{{\rm day}}

\begin{document}

\title{Optimal Microlensing Observations}

\author{Andrew Gould\altaffilmark{1}\altaffiltext{1}
{Alfred P.\ Sloan Foundation Fellow}}
\affil{Ohio State University, Department of Astronomy, 
174 West 18th Ave., Columbus, OH 43210}
\affil{E-mail: gould@astronomy.ohio-state.edu}

\begin{abstract}

	One of the major limitations of microlensing observations toward the 
Large Magellanic Cloud (LMC) is the low rate of event detection.  What can
be done to improve this rate?  Is it better to invest telescope time in
more frequent observations of the inner high surface-brightness fields, or 
in covering new, less populated outer fields?  How would a factor 2 improvement
in CCD sensitivity affect the detection efficiency?  Would a
series of major (factor 2--4) upgrades in telescope aperture, seeing, sky
brightness, camera size, and detector efficiency increase the event rate
by a huge factor, or only marginally?  I develop a simplified framework to
address these questions.  With observational resources fixed at the level of
the MACHO and EROS experiments, the
biggest improvement (factor $\sim 2$) would come by reducing the time spent
on the inner $\sim 25\,\rm deg^2$ and applying it to the outer
$\sim 100\,\rm deg^2$.  By combining
this change with the characteristics of a good medium-size telescope
(2.5 m mirror, $1''$ point spread function, thinned CCD chips, 
$1\,\deg^2$ camera, and dark sky), it should be possible to increase the
detection of LMC events to more than 100 per year (assuming current estimates
of the optical depth apply to the entire LMC).

\keywords{dark matter -- Galaxy: halo -- gravitational lensing 
-- Magellanic Clouds}
\end{abstract}

\section{Introduction}

	Microlensing observations toward the Large Magellanic Cloud 
(LMC) have yielded puzzling results:  The event rate toward the LMC
is much too high to be caused by known populations of stars but the 
$M\sim 0.4\,M_\odot$ mass of the lenses (as inferred from the 
$t_\E\sim 40\,$day time scale of the events) is too heavy to be due to a
halo of
brown dwarfs (Alcock et al.\ 1997a; Ansari et al.\ 1997a).  Moreover,
if the LMC events were due to halo lenses, one would expect similar events
toward the Small Magellanic Cloud (SMC).  However, both of the events
detected to date toward the SMC show signs of being SMC self-lensing
(Palanque-Delabrouille et al.\ 1998; Afonso et al.\ 1998; Alcock et al.\ 1997b,
1998; Albrow et al.\ 1998).

	The two most difficult obstacles to unraveling the nature of the
lenses are the low overall event detection rate and the lack of information 
about individual events.  The first two years of LMC observations by the MACHO
collaboration yielded only 8 candidate events over the inner $11\,\deg^2$
(Alcock et al.\ 1997a), making it difficult to discern non-uniformities
in the spatial distribution of the events as one would expect if they were
predominantly due to LMC self-lensing.  For most events, the only information
is the time scale $t_\E$ which is a complicated combination of the three
quantities one would like to know about the lens, its mass $M$, distance 
$\dol$, and transverse speed $v$ relative to the observer-source line of 
sight,
\beq
t_\E = {r_\E\over v},\qquad r_\E^2 = {4 G M \dol\dls\over c^2\dos}.
\label{eqn:tedef}
\eeq
Here $r_E$ is the Einstein radius, and $\dol$, $\dls$, and $\dos$, are
the distances between the observer, lens, and source.  Hence, for most
events one cannot tell how far the lens is or how fast it is going, 
characteristics which, if known, would clearly distinguish between the 
halo-lens and LMC-lens hypotheses.

	The low overall event detection rate exacerbates the problem of
lack of information about individual events.  For a small
fraction of events it is possible to extract additional information.
For example, if the lens is a binary and the source crosses the caustic
in the binary-lens magnification pattern, then one can measure the time
it takes for the lens to cross the source angular radius (known from its
color, flux, and the Planck law) and so determine the proper motion of the 
lens.  In fact, this was how one of the two SMC events was inferred to be
self-lensing.  If the event is sufficiently long, the reflex motion of the
Earth gives rise to parallax distortions of the light curve (Gould 1992),
and if the source is bright enough to allow measurement of this subtle
effect, then one can measure the combination $\tilde v = (\dos/\dls)v$
which is very different for LMC and halo lenses.  This is how the other
SMC event was inferred to be self-lensing.  Another rare (and not yet
definitively observed) effect which can yield a proper motion measurement
is a binary source (Han \& Gould 1997).  In the future, it may be possible
to measure parallaxes and/or proper motions using the 
{\it Space Interferometry Mission} (Boden, Shao, \& Van Buren 1998)
or the {\it Space Infrared Telescope Facility} (Gould 1999), but only for
sufficiently bright ($V\la 20$) and hence relatively rare sources.  
If the detection rate could be improved, the total number of events for which
more information could be extracted would likewise increase.

	It is reasonable to suppose that given larger telescopes, larger and
more efficient detectors, smaller point spread functions (PSFs), fainter sky, 
and better weather, it would be possible to increase the event detection rate.
But by how much?  Would a massive series of upgrades be worth the effort
and expense?  After seven years of microlensing experiments, there are no
published works that address this question.  It is not even known, for example,
whether it is better to spend the telescope resources presently available
intensively observing the brighter regions of the LMC where there are more
sources, or observing these less intensively and applying the telescope time
so saved to the outer regions of the LMC with lower surface brightness.

	One reason for the slow progress on this front is that the
problem of estimating the detection rate for a given set of observational
parameters (often called the ``efficiency'') is very time consuming.
For example, after many years of effort, the MACHO collaboration has
only recently succeeded in developing a pipeline that takes an arbitrary
series of observations and returns an efficiency estimate (K.\ Griest 1998,
private communication).  To actually apply this algorithm to the accumulated
data set will require many months of computer time.  Hence, the determination
of efficiencies for a multiplicity of hypothetical observing programs seems
like an intractable problem.

	Estimating the real efficiencies is complicated because
the real detection algorithms are complicated.  These require the formation of
a template image and the identification on the template 
of a set of ``stars''.  The number
of such ``stars'' is limited by the number of resolution elements in the
template image, but each ``star'' may be composed of several real stars
whose light is all blended together.  Whether lensing of one of these stars
is detectable depends on the combination of other stars in and near the
resolution element as well as on the temporal pattern and intensity of the
observations.

	However, for purposes of understanding the {\it relative} efficiency
of different observational strategies, these details of the detection
algorithm are not important: their effects approximately cancel when one
compares one strategy with another.  Moreover, present PSF-fitting 
detection algorithms 
are likely to be replaced in the future by pixel lensing (image subtraction)
techniques.  In contrast to PSF fitting, the mathematical description of
pixel lensing is extremely simple (Gould 1996).  
Hence, by using the pixel-lensing formalism
one can understand the whole range of possible observation strategies in 
terms of a few easily understood parameters.  Although the absolute number
of events detected by current algorithms will be overestimated by the
pixel-lensing formalism, this overestimate is not likely to be more than
a factor of 2.  See \S\ 5.2. 
More importantly, the relative number for different strategies
should be quite accurate.

\section{Pixel-Lensing Formalism}

In the standard microlensing formalism, one imagines that one is monitoring
an isolated star of unmagnified flux $F_0$ and that it is magnified by a lens 
to a flux (Paczy\'nski 1986)
\beq 
F(t;t_0,\beta,t_\E,F_0) = F_0 A[u(t;t_0,\beta,t_\E)],\qquad u(t) = 
\biggl[{(t-t_0)^2\over t_E^2} +\beta^2\biggr]^{1/2},
\label{eqn:foft}
\eeq
where $t_0$ is the time of maximum magnification, $\beta$ is the impact
parameter in units of $r_\E$, and $A(u)$ is the magnification,
\beq
A(u) = {u^2 + 2\over u(u^2+4)^{1/2}}.
\label{eqn:aofu}
\eeq
Actually, in crowded fields one can never assume that the source star is
truly isolated.  In fact, even isolated stars can have luminous binary 
companions or the lens could be luminous.  Hence one must generally write
equation (\ref{eqn:foft}) as
$F(t;t_0,\beta,t_\E,F_0,B) = F_0 A[u(t)] + B$, where $B$ is the sum total of
all unlensed sources in the aperture.  This can in turn be rewritten,
\beq
F(t;t_0,\beta,t_\E,F_0,\tilde B) =F_0 \{A[u(t;t_0,\beta,t_\E)] -1\} + \tilde B,
\label{eqn:fofttilde}
\eeq
 where $\tilde B=F_0 +B$ is the baseline flux.  Since the baseline flux
is ordinarily well measured by the numerous observations away from the event,
it can easily be subtracted from the remaining flux measurements.  Hence,
equation (\ref{eqn:fofttilde}) can effectively be rewritten,
\beq
\tilde F(t) = F(t)-\tilde B =F_0 \{A[u(t)] -1\}.
\label{eqn:foftt}
\eeq
Equation (\ref{eqn:foftt}) was originally written to describe lensing
toward M31, not the LMC.  For M31, one does not begin with the delusion that
one is monitoring an isolated star because the field
contains virtually no resolved stars.  Rather, one recognizes that the only
observable quantity is the difference in flux $\tilde F(t)$ between the 
present epoch and the baseline (Crotts 1992; Baillon et al.\ 1993).  
Consider a single observation with exposure time $\texp$ by a telescope
that records $\alpha$ electrons per unit flux per unit time.  Then the 
signal-to-noise ratio, $Q$, of the observation is
\beq
Q(t) = {F_0\{A[u(t)] -1\}\alpha \texp\over 
[\{S_t\Omega_\psf + F_0 A[u(t)]\}\alpha \texp]^{1/2}}.
\label{eqn:qdef}
\eeq
where $\Omega_\psf$ is the angular area of the PSF and 
$S_t\Omega_\psf$ is the total flux (including neighboring stars plus sky)
inside the aperture.  For M31, the surface brightness is sufficiently
uniform that $S_t$ can be taken to be the average surface brightness near
the source.  For the LMC, this approximation no longer holds: sometimes
the galaxy light falling into the aperture will be significantly more
than average and sometimes less.  I will assume that for the purpose of 
estimating efficiencies, these variations cancel out, and I adopt equation
(\ref{eqn:qdef}) as written.  Suppose that a series of observations are
made roughly uniformly over the event, between times $t_-$ and $t_+$,
with a mean exposure time per day $\texp$, always with the same seeing.
Then $\Delta \chi^2$, the square of the total signal-to-noise ratio
is given by
\beq
\Delta \chi^2 = \sum_i Q^2(t_i)
= \alpha \texp{t_e\over \day}G(S_t\Omega_\psf,F_0,\beta,\tau_\pm),
\label{eqn:deltachi}
\eeq
where,
\beq
G(F_s,F_0,\beta,\tau_\pm)=
\int_{\tau_-}^{\tau_+}d\tau
{F_0^2 \{A[u(\tau,\beta)] - 1\}^2\over F_s + F_0 A[u(\tau,\beta)]},
\label{eqn:gdef}
\eeq
and where $\tau \equiv (t -t_0)/t_\E$, and $F_s\equiv S_t\Omega_\psf$.
	For $|\tau_\pm|\ga 1.5$, G is only weakly dependent on $\tau_\pm$.
For simplicity, I will henceforth adopt $\tau_\pm = \pm 2$ and remove 
$\tau_\pm$ as arguments of $G$.

\section{Luminosity Function}

	I construct a luminosity function (LF) from the observed apparent $R$
band LF of the MACHO collaboration (D.\ Alves 1998, private communication)
and the absolute $M_V$ band LF of Holtzman et al.\ (1997) derived from 
{\it Hubble Space Telescope (HST)} data.  For the latter,
I first recover the observed $V$ band LF by adding the distance modulus of the
LMC ($\mu_\lmc=18.5$) and the extinction ($E(B-V)=0.1$) 
adopted by the authors.  I then convert to $R$ band
using the relation $V-R = (M_V-2.89)/6.74$.  This is actually valid only for 
main-sequence stars, but these are the vast majority of the {\it HST} stars,
and in any event the $V$-to-$R$ conversion has almost no impact on the
results.  The two LFs are shown in Figure \ref{fig:one}.  The MACHO data become
incomplete for $R\ga 20$.  The {\it HST} data suffer from small number 
statistics for $R\la 19$.  I therefore match the two by eye in the overlap
region (as indicated in Fig.\ \ref{fig:one}) 
and construct the final LF by using
MACHO for $R\leq 20$ and {\it HST} for $R>20$.  Note that the {\it HST} LF
itself suffers from serious incompleteness for $R>26$.  However, this has
almost no impact on the present study since these fainter stars contribute
very little to the total light (and so to the normalization of the LF) and
even less to observable microlensing events.  The LF in Figure \ref{fig:one}
is normalized to a total flux $F_*$ corresponding to $R=3.85$.  
This is the integrated light in 
$1\,\deg^2$ assuming 10 times the unit surface brightness arbitrarily 
adopted by de Vaucouleurs (1957) for his surface brightness map of the LMC.  
A region with $F_*\,\deg^{-2}$ has a surface brightness of $R=21.63$ which
is typical of the inner $10\,\deg^2$ of the LMC.
I will therefore use this unit of integrated flux throughout this paper,
\beq
{F\over F_*} = 10^{-0.4(R-3.85)}.
\label{eqn:fstar}
\eeq

\section{Event Detection Functions}

	I now suppose that all events with $\Delta\chi^2$ greater than some
minimum $\Delta \chi^2_\min$ are detected.  For each star of flux $F_0$,
and impact parameter $\beta$,
one can therefore define a minimum exposure time (per day) required
for detection of the event (see eq.\ \ref{eqn:deltachi})
\beq
t_{\rm exp} = {\Delta \chi^2_\min\over\alpha (t_\E/\day)G(F_s,F_0,\beta)}.
\label{eqn:tmineval}
\eeq
I now assume that $\Delta \chi^2_\min$, $\alpha$ and $t_\E$ are all fixed.
 For $\Delta\chi^2_\min$, I adopt the value
used by the MACHO collaboration in their two-year LMC study, 
$\Delta \chi^2_\min=500$ (Alcock et al.\ 1997a).  I adopt $t_\E=40\,$days,
the typical time scale measured by MACHO (Alcock et al.\ 1997a).  Of course,
the actual observed values of $t_\E$ cover a broad range of a factor $\sim 8$.
However, I show below that this simplifying assumption has almost no impact
on the results.  I adopt 
$\alpha=125\,\rm s^{-1}$ at $R=20$, corresponding to what is expected from a 
2.5 m telescope with a thinned CCD and standard Cousins $R$ filter.  
These are the characteristics of the ``next generation'' microlensing
experiment proposed by C.\ Stubbs (private communication).  I will consistently
use the ``next generation'' characteristics in my initial example.
After
fixing these parameters, $t_{\rm exp}$ is a function only of 
$F_s$, $F_0$, and $\beta$.  I then integrate over the LF and a uniform
distribution in $\beta$ to obtain the event rate as a function
of the minimum daily exposure time necessary to observe them,
\beq
{d \Gamma_i\over d t_{\rm exp}} = {2\over\pi}\,{\tau\over t_\E}\,
{S_i\Omega_\ccd\over F_*}\int_0^{0.66}d\beta\int d F_0 \Phi(F_0) \delta 
\biggl[t_{\rm exp} - 
{\Delta \chi^2_\min\over\alpha (t_\E/\day)G(F_s,F_0,\beta)}\biggr]
\label{eqn:dndt}
\eeq
where $\delta$ is the Dirac $\delta$-function, $\Phi$ is the LF normalized
to $F_*$
(see Fig.\ \ref{fig:one}), 
$S_i$ is the surface brightness of field $i$, and $\Omega_\ccd$ is the area
of the CCD.
I assume an optical depth $\tau=2.9\times 10^{-7}$, the
best-fit value for the MACHO two-year study (Alcock et al.\ 1997a). 
Note that I have restricted the integration to 
$\beta\leq 0.66$, corresponding to a minimum peak magnification $A_{\rm peak}
\geq 1.75$, again following the MACHO selection criteria 
(Alcock et al.\ 1997a).  Figure \ref{fig:two} shows the normalized 
cumulative distribution function $(F_*/S_i\Omega_\ccd)\Gamma_i(t_{\rm exp})$ 
where $\Gamma(t_{\rm exp})$ 
is the integral of equation (\ref{eqn:dndt}), 
assuming a 180-day observing season per year.  
Five different values of LMC surface brightness are shown
ranging from $S_\lmc= 2.3 F_*\,\deg^{-2}$ ($R=20.73\,\rm mag\,\rm arcsec^{-2}$)
characteristic of the LMC bar to
$S_\lmc= 0.12 F_*\,\deg^{-2}$ ($R=23.93\,\rm mag\,arcsec^{-2}$)
characteristic of the region $\sim 5^\circ$ from the LMC center.  For each
of these calculations, I have assumed a sky $S_{\rm sky}$ of 
$R=21.0\,\rm mag\,arcsec^{-2}$, and a PSF size
$\Omega_\psf=\pi \,\rm arcsec^2$.

	Most of the conclusions of this paper can be extracted from a
careful inspection of Figure \ref{fig:two}.  First, the five
curves look very similar, differing by only $\sim 18\%$ at the canonical
exposure time $\texp=5\,$minutes.  This means that, for fixed
exposure time, the number of detectable events is essentially proportional
to the surface brightness (which has been factored out of Fig.\ \ref{fig:two}).
Second, the slope of these curves at $\texp=5\,$minutes is
$d\ln \Gamma/d\ln \texp \sim 0.23$.  That is, a factor 2 increase
in exposure time increases the rate of event detection by only $\sim 16\%$.
Hence, faced with the choice of doubling the exposure time on a high
surface-brightness field or observing a new field with 1/5 the surface 
brightness, one should choose the latter.  In fact, I will show in \S\ 5,
that essentially the whole LMC should be monitored.  Third, the event rate for
the canonical $\texp=5\,$minute exposures is surprisingly high,
$\sim 5\,F_*^{-1}$ events per year.  Since the total flux from the LMC
is $\sim 36\,F_*$, this implies that over 100 events per year could be detected
if the survey covered the whole LMC.

	Figure \ref{fig:two} also allows one to understand why using 
using the average event time scale is adequate for predicting the total event
rate.  From equation (\ref{eqn:deltachi}), it follows that
events that are a factor of two shorter than average suffer the same loss
of signal-to-noise ratio as events with half the exposure time.  Hence, they
suffer the same loss of detection rate, i.e., 23\%.  This means that over the
entire factor $\sim 8$ range of observed time scales, there is only a few
tens of percent difference in detection rate.  Thus, the detection rate for
the mean time scale is an excellent proxy for the mean detection rate.

\section{Optimal Strategies}

	The formalism developed in the previous section can be used to
estimate the event detection rate for various observational programs and
to optimize detection efficiency for a given set of equipment.
I first analyze the ``next generation'' experiment (whose characteristics
are reflected in Fig.\ \ref{fig:two}) and then compare this with the 
current MACHO (Alcock et al.\ 1997a) and EROS (Ansari et al.\ 1997a) 
experiments.

\subsection{Next Generation}

	As described in \S\ 4, the ``next generation'' experiment proposed
by C.\ Stubbs (1998, private communication) would have a 2.5 m telescope,
an $\Omega_\ccd=1\,\deg^2$ 
camera with thinned CCDs (and so $\alpha=125\,e^-\,\rm s^{-1}$
at $R=20$), a dark sky ($R=21.0\,\rm mag\,arcsec^{-2}$), and a small PSF 
($\Omega_\psf=\pi \,\rm arcsec^2$).

	Let $\Gamma_i(S_i,S_\sky,\texp)$ be the event rate
for a $1\,\deg^2$ field with surface brightness $S_i$, and background flux
$F_s = \Omega_\psf (S_i + S_\sky)$.  The total rate of detectable
events is then
\beq
\Gamma_\tot = \sum_i \Gamma_i(S_i,S_\sky,\texp).
\label{eqn:ntot}
\eeq
I maximize $\Gamma_\tot$ subject to a constraint on the total amount of 
observing
time.  I assume an average of 6.5 hours per night are available for
observations over a 180 day observing season and that 49 minutes
of this time are lost to overhead (readout and pointing).  I discuss this
figure further below.  I assume that 20\% of the time is lost to weather
and 25\% of the remaining time is lost to (or at any rate degraded by)
the Moon.  I construct an $11^\circ\times 11^\circ$ 
surface-brightness grid using the
de Vaucouleurs (1957) map.  I find a total event rate 
$\Gamma_\tot=129\,\yr^{-1}$, with a distribution of exposure times shown in
Figure \ref{fig:three}.  Note that the exposure times are roughly proportional
to surface brightness.  This can be understood from Figure \ref{fig:two}:
if the curves were exactly straight lines and were completely independent
of surface brightness, then the proportionality would be exact.  That is,
we would have $\Gamma_i = C_1 S_i\ln(C_2\texp)$, so that 
$d\Gamma_i/d\texp = C_1 C_2 S_i/\texp$ where $C_1$ and $C_2$ are constants.
Detection is maximized when these derivatives are equal in all fields, which
occurs if $\texp\propto S_i$.  The inner fields are typically observed for
about 5 minutes, while the outer fields are typically observed for 1 minute
or less.  I assume that there is 1 minute of telescope overhead time per
exposure, so these very short exposures in the outer fields seem wasteful.
I therefore assume that the 
inner $25\,\deg^2$ are observed every available night
while the outer $96\,\deg^2$ are observed only every third night.  This
schedule accounts for my estimate of 49 minutes of overhead per night.

	The total event rate is actually not very sensitive to
the exact observation strategy, {\it provided that the whole LMC is observed}.
For the optimal exposure time, there are 129.0 events per year, 69.5 in the
inner $25\,\deg^2$ and 59.5 in the outer $96\,\deg^2$.  However, if exposure
times are set to be equal, I find 127.3 events with 67.3 in the inner fields
and 60.0 in the outer fields.  On the other hand, if only the inner 
$25\,\deg^2$ is observed (and the overhead time is consequently cut to 25 
minutes), then only 70.9 events are expected.  This confirms the conclusion
I drew from inspection of Figure \ref{fig:two} that additional telescope
time is better spent on low-surface-brightness fields than on intensive
monitoring of the inner fields.

	Clearly, however, when the surface brightness falls sufficiently low,
it must be counter-productive to observe a field.  To determine that point, I
return to the optimal solution.  As noted above, $d\Gamma_i/d\texp$ must
be equal in all fields, and its value is 0.18 (events/year)/minute.  Since, the
overhead for the outer fields is 1/3 minute, this implies that observation
of a field is counter-productive if the event
rate falls below 0.06 events/year.  I find that this occurs at
$S\Omega_\ccd = 0.02F_*$ ($R=25.9\,\rm mag\,arcsec^{-2}$) which is generally
fainter than the inner $121\,\deg^2$.

\subsection{MACHO Experiment}

	I now apply the same formalism to the MACHO experiment (Alcock et al.\
1997a).  I assume $\alpha=31\,e^-\,\rm s^{-1}$ at $R=20$ corresponding to
a 1.25 m telescope with unthinned CCDs but a broader $R$ passband.  I assume
a brighter sky ($R=19.5\,\rm mag\,arcsec^{-2}$), a larger PSF 
$(\Omega_\psf= 4\pi\,\rm arcsec^2)$, and a smaller camera $\Omega_\ccd=
0.5\,\deg^2$.  I assume a 50\% time loss to weather, but only 15\% to the moon
(because the sky is already so bright).  I continue to assume 1 minute of 
overhead per exposure.  I then find a total of 27.6 events/year, or 15.8 if
observations are restricted to the inner $25\,\deg^2$.

	As a consistency check, it is important to try to make contact with 
the two-year MACHO results based on an inner region of $11\,\deg^2$.  Eight
candidate events were detected.  Recall that I normalized 
the event rate to the optical depth $(\tau = 2.9\times 10^{-7})$ estimated
by MACHO based on these eight events.  Since MACHO spent substantial time
observing other regions of the LMC (even though they only reported on these
$11\,\deg^2$) I mimic the MACHO observations by assuming that the inner
$25\,\deg^2$ were monitored, but count events only for the brightest
$11\,\deg^2$.  I then find 10.6 events per
year, substantially more than the four
events per year actually observed.  

	Part of the difference is undoubtedly due
to the fact that I have assumed a pixel-lensing analysis, while MACHO carried
out a Dophot-based analysis.  Any unresolved stars that happened to lie
within the PSF of a template ``star'' will be effectively monitored and so
subject to detection in a Dophot-based analysis.  However, lensing of 
unresolved stars lying between template ``stars'' will be missed.
Melchior et al.\ (1998) also concluded that a pixel-lensing analysis of LMC
observations would increase the event detection rate substantially.  On the
other hand,
it is possible that part of the difference between the 10.6 events predicted
in my analysis and the 4 observed by MACHO is that my simplified analysis 
fails to reflect real
effects that would diminish the effectiveness of both a Dophot-based and a
pixel-lensing analysis.  (Note, however, that Poisson statistics is {\it not}
a possible cause since my analysis was normalized to the optical depth
based on the four events actually detected.)  To the extent that the
difference between the 10.6 events predicted and 4 events observed is due
factors that are common to a pixel-lensing and Dophot-based analyses, my 
estimates of the event rate in a ``next generation'' experiment should also
be scaled down.

\subsection{EROS experiment}

	For the EROS II experiment,
I assume $\alpha=20\,e^-\,\rm s^{-1}$ at $R=20$ corresponding to
a 1 m telescope with unthinned CCDs,
$\Omega_\psf= 4\pi\,\rm arcsec^2$, and a camera size 
$\Omega_\ccd=1\,\deg^2$.  I assume sky, weather, and moon conditions
similar to the ``next generation'' parameters, and telescope overhead of
2.5 minutes.  I find a total of 46.1 events/year, of which 24.7 are in the
inner $25\,\deg^2$, or a total of 25.5 if
observations are restricted to the inner $25\,\deg^2$.  EROS
carries out a number of non-microlensing projects which reduce the time
available for LMC observations, so these rates may be slightly overestimated.

\section{Toward a Pixel Analysis of the LMC}

	As I indicated in \S\ 5.2, of order half the events predicted by
my analysis are not being found by Dophot type analyses, and could only
be found using pixel lensing.  Here I review the progress being made
toward a pixel analysis of the LMC and offer some ideas on how to overcome
the remaining obstacles.

	Image subtraction has been applied in three density domains.  In
order of increasing stellar density, these are
1) planetary nebula and supernova 
searches in high-latitude fields where all stars are isolated,
2) pixel lensing searches of the LMC, SMC, and bulge which are crowded fields 
of {\it resolved} stars, and
3) pixel lensing searches of M31 where the stars are {\it unresolved}.

Three substantially different ideas have emerged on how to carry out the
analysis.  Ciardullo, Tamblyn, \& Phillips (1990) and
Phillips \& Davis (1995) convolve the better-seeing image $(R)$ to the 
resolution of the worse seeing image $(I)$ and then subtract the two.  
The kernel of the convolution $(\Psi)$
is determined by dividing the Fourier coefficients of the PSFs of the
two images as measured from the isolated stars (and a prescription is given
for suppressing high-frequency noise).  
Any star
that has changed brightness between the two exposures should then appear
as an isolated PSF on the difference image, $D = I-\Psi\otimes R$.  
This method was devised for the
lowest density regime (1) but was taken over essentially unchanged by
Tomaney \& Crotts (1996) for use in the highest density regime (3).
A.\ Tomaney (1997, private communication) has been applying this technique
to MACHO bulge and LMC data, i.e. the intermediate regime (2).
Alard \& Lupton (1998) developed a substantially different method for finding
the convolution kernel.  They simply write the kernel as a linear combination
of basis functions $\Psi = \sum_i a_i F_i$, and determine the coefficients
by writing $\chi^2 = \sum_{x,y} [I(x,y) -\sum_i a_i f_i(x,y)]^2/\sigma(x,y)^2$
where $f_i\equiv F_i\otimes R$ and $\sigma(x,y)^2$ is the variance of the 
difference at pixel $(x,y)$.  The coefficients can be found using standard
linear techniques.  Alard \& Lupton (1998) applied this to the bulge.
EROS (Afonso et al.\ 1998) has adopted this method to
do photometry of events, but still uses PSF-fitting based techniques to
find the events.  Finally, Ansari et al.\ (1997b) developed yet another
technique: direct subtraction of pixels, without convolution but with a
seeing correction.  Melchior et al.\ (1998) have applied this method
to a subset of EROS data toward the LMC.

	What is the best approach for finding events in the LMC?
Alard \& Lupton (1998) argue (correctly, I believe, in the case of the LMC) 
that their method uses all the information and so is optimal.  (In the case
of M31 and high-latitude fields, virtually all the information about the
PSF is in the handful of isolated field stars and the rest of the image
contains only noise, so the Phillips \& Davis 1995 method is probably the 
best.)\ \ 
However, EROS finds that photometry of a single source requires about 1 minute
per image (E.\ Aubourg 1998, private communication) so that applying this
method to a search over $100\,\deg^2$ would require formidable computer
resources.  It may be possible to apply the Phillips \& Davis (1995) method
as a search technique to the LMC, but it is unlikely to produce as high
quality photometry as Alard \& Lupton (1998) because there are few if
any isolated stars on which to measure the PSF.  
Thus, once an event is found (by whatever means),
the Alard \& Lupton method should be used for photometry 
(as EROS is now doing).  The Ansari et al.\ (1997b) method of simple 
pixel subtraction yields substantially worse photometry than either
of the image convolution techniques.  Nevertheless, it is certainly adequate
to find the events ($\Delta \chi^2>500$) discussed in this paper.  It is
also computationally straight forward.  Hence, a useful approach would be
to apply this method as a loose filter to locate candidate events and then
use Alard \& Lupton (1998) photometry to make a final determination of
the status of the event and to measure its parameters if it is confirmed
to be microlensing.  In fact, with the superior photometry of
Alard \& Lupton (1998) on all events, it might be possible to push the
detection threshold below the current minimum, $\Delta\chi^2_\min=500$.

{\bf Acknowledgements}:  
I thank C.\ Stubbs for his persistent prodding, A.\ Becker for help
determining the LMC luminosity function, and B.S.\ Gaudi for a careful
reading of the manuscript.
This work was supported in part by grant AST 97-27520 from the NSF.

\clearpage

\begin{figure}
\caption[junk]{\label{fig:one}
Luminosity functions (LFs) for the LMC normalized so that the integrated flux
is $R=3.85$, i.e., 10 de Vaucouleurs (1957) surface-brightness units 
integrated over
$1\,\deg^2$.  The complete LF is constructed from the
MACHO LF ({\it bold}) (D.\ Alves 1998, private 
communication) for $R\leq 20$, and the {\it HST} LF ({\it solid})
(Holtzman et al.\ 1997) for $R>20$.  The {\it HST} LF is first transformed
from $V$ to $R$ band before being plotted here.  The relative normalization
between the two LFs is set from the overlap region $19<R<20$.
}
\end{figure}

\begin{figure}
\caption[junk]{\label{fig:two}
Event rate per year per $F_*$ of LMC flux, as a function of exposure time per 
day.  Here $F_*$ (eq.\ \ref{eqn:fstar}) is the flux corresponding to
$R=3.85$ (which is typical of the flux from $1\,\deg^2$ in the central
$10\,\deg^2$ of the LMC).  
To obtain the true rate, multiply by $S_i\Omega_\ccd/F_*$.
Shown (left to right) are curves for surface brightness 0.12, 0.41, 1.2,
1.5, and 2.3 $F_*\,\deg^{-2}$, corresponding to a range of $R=23.93$ to
$R=20.73\,\rm mag\,arcsec^{-2}$.  Characteristics of the ``next generation''
microlensing experiment have been assumed: 2.5 m telescope with thinned
CCDs, sky brightness of
$R=21.0\,\rm mag\,arcsec^{-2}$, PSF size $\Omega_\psf=\pi\,\rm arcsec^2$.
}
\end{figure}

\begin{figure}
\caption[junk]{\label{fig:three}
Optimal distribution of exposure times for 121 $1\,\deg^2$ LMC fields. 
Exposure times are chosen
to maximize the total number of events assuming uniform optical depth across
the LMC.  The abscissa is the surface brightness $S$ in units of
$R=21.63\,\rm mag\,arcsec$.
Assumptions are the same as in Fig.\ \ref{fig:two}.  The optimal exposure
time is almost exactly proportional to $S$.
However, equal exposure times in all fields reduces the total number
of events by only about 2\%.
}
\end{figure}

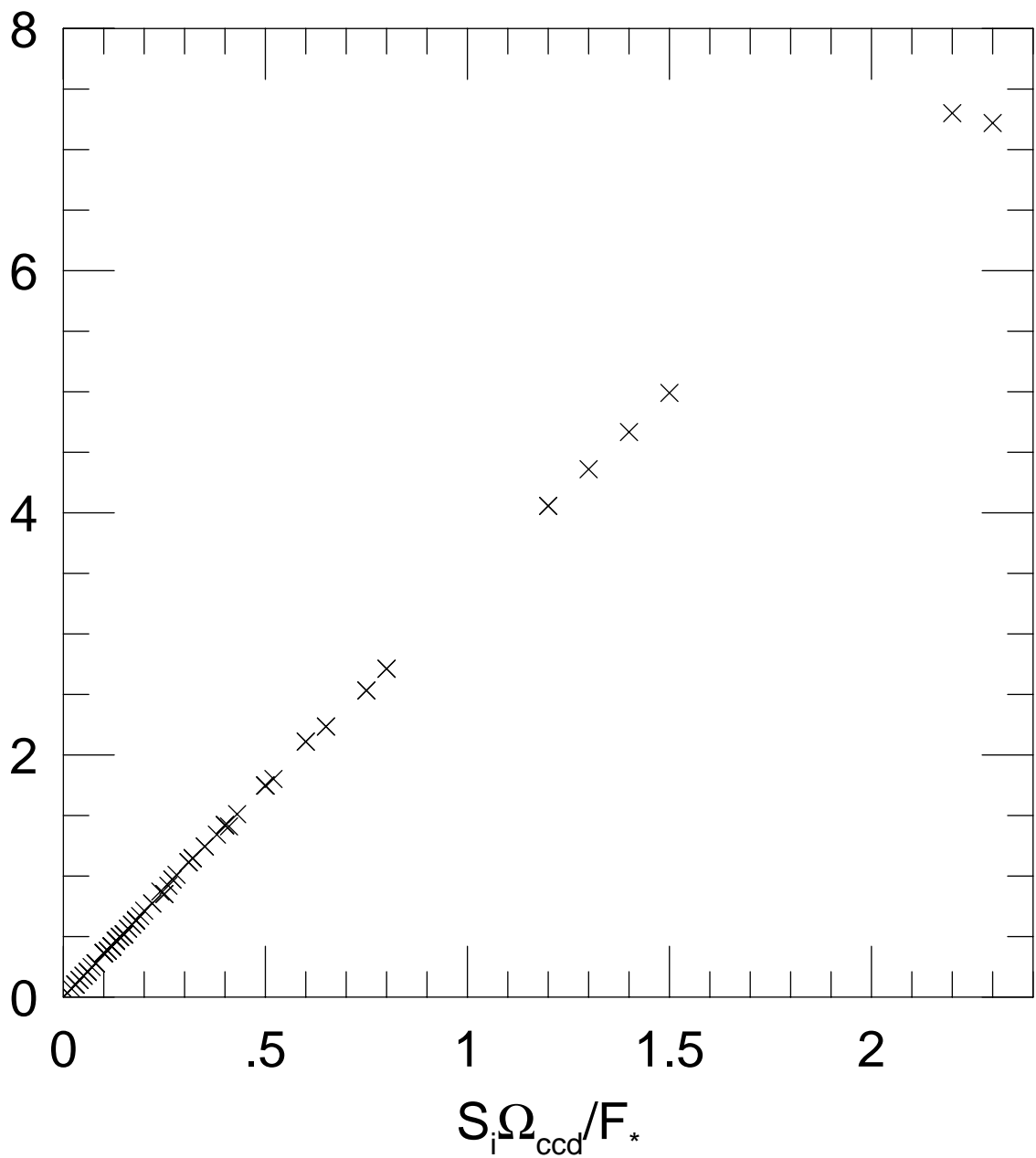
\end{document}